\documentclass[12pt,a4paper]{article}

\usepackage[utf8]{inputenc}
\usepackage[spanish]{babel}
\usepackage[left=2.5cm,top=2cm,right=2.5cm,bottom=2cm]{geometry} 
\usepackage{titlesec}
\usepackage{hyperref}
\usepackage{amssymb}
\usepackage{amsmath}
\usepackage{amsthm}
\usepackage{array}
\usepackage{graphicx}
\usepackage{subfig}
\usepackage{xcolor}
\usepackage[all]{xy}
\usepackage{comment}
\usepackage{algpseudocode}
\usepackage{standalone}
\usepackage{tikz}

\title{Uso de herramientas digitales matemáticas en la Educación Secundaria}

\author{José Manuel Fernández-Barroso\\
	\small{Department of Mathematics,}\\
	\small{University of Extremadura,} \\
	\small{Badajoz, Spain}\\
	\small{ferbar@unex.es}\footnote{Afiliación durante la realización del trabajo: UNED}}
\date{}

\begin{document}
	
	\maketitle

\begin{abstract}
Las Tecnologías de la Información y la Comunicación (TIC) están cada día más presentes en nuestra sociedad y por tanto en el ámbito educativo. En apenas dos décadas hemos pasado de una enseñanza basada, en muchos casos, en las clases magistrales a una enseñanza en la que metodologías como el aula invertida o la gamificación tienen más fuerza que nunca.

A lo largo del trabajo hemos realizado una encuesta a docentes y alumnado con el objetivo de comparar los conocimientos en herramientas digitales, su uso y su aceptación. Hemos utilizado WxMaxima y Geogebra como herramientas didácticas. Para ello propondremos un ejercicio de la \textit{Evaluación de Bachillerato para el Acceso a la Universidad} (EBAU) relacionado con la geometría, analizando sus puntos fuertes y débiles en comparación con la resolución manual. Por útlimo, expondremos algunas conclusiones y posibles líneas de investigación acerca de las herramientas digitales, así como una propuesta de curso introductorio a WxMaxima y Geogebra con el que formar al docente de secundaria.
\end{abstract}
\begin{quote}
\begin{center}
\textbf{Abstract}
\end{center}
Information and Community Technologies (ICT) are very present in our society nowadays and particularly in the educative field. In just two decades, we have passed from a learning based, in many cases, on the master lessons to one such that methodologies like the flipped classroom or the gamification are stronger than ever.

Along this work, we have done a study to teachers and students with the main objective to compare the knowledge on digital tools, their use and their acceptation. We use WxMaxima and Geogebra in order to solve an exercise of \textit{Evaluación de Bachillerato para el Acceso a la Universidad} (EBAU) related with Geometry, comparing their ins and outs with the manual solution. Finally, we expose some conclusions and some possible research lines about digital tools, as well as a proposition of an introductory course on WxMaxima and Geogebra in order to teach the teachers.
\end{quote}

	\textbf{Keywords:} TIC; Geogebra; WxMaxima; Herramientas digitales; Educación Matemática; Competencia Digital.
	
	\textbf{MSC2020:}
	97D10; 97N80; 97U10.

\section{Introducción}

La \textit{Ley Orgánica 2/2006, de 3 de mayo, de Educación}, LOE en adelante, define las competencias como ``las capacidades para aplicar de forma integrada los contenidos propios de cada enseñanza y etapa educativa, con el fin de lograr la realización adecuada de actividades y la resolución eficaz de problemas complejos''. Todas las leyes que han sucedido a esta, han apostado por un carácter competencial y cada vez tienen más fuerza dentro de la enseñanza. En la vigente \textit{Ley Orgánica 3/2020, de 29 de diciembre, por la que se modifica la Ley Orgánica 2/2006, de 3 de mayo, de Educación} (LOMLOE en adelante), podemos encontrar la ``Competencia Matemática y de ciencias, tecnología e ingeniería'' (STEM) y la ``Competencia digital''. Según el Ministerio de Educación y Formación Profesional, la competencia matemática ``implica la capacidad de aplicar el razonamiento matemático y sus herramientas para describir, interpretar y predecir distintos fenómenos en su contexto'', mientras que la competencia digital implica el uso creativo y crítico de las tecnologías de la información y la comunicación, siempre de manera segura y controlada, por lo que no sólo implicará destrezas en el acceso o procesamiento de la información, sino también del filtrado y comprobación de la veracidad de dicha información.

Es evidente la relación directa entre las matemáticas y el uso de las tecnologías, ya que los nuevos avances científicos permiten, por ejemplo, el mejor cifrado y encriptado de contraseñas en internet, pero la relación entre informática y matemáticas parece más difusa. Esta relación se propicia cuando, para descubrir nuevos resultados matemáticos, se hace uso de las herramientas informáticas para realizar cálculos que hacerlos manualmente nos llevarían una eternidad. Si echamos la vista atrás, no hace mucho los logaritmos se calculaban mediante interpolaciones a partir de valores ya calculados en unas tablas. Estas tablas quedaron obsoletas cuando aparecieron las calculadoras. Estamos quizás ante la siguiente brecha digital, donde los ordenadores han venido a sustituir a las calculadoras, como estas hicieron con las tablas de logaritmos y otros muchos cálculos manuales. No podemos dar el siguiente paso tecnológico a los ordenadores, tablets o pizarras digitales si el personal docente no tiene suficientes conocimientos sobre su utilización.

Es evidente entonces, que la incorporación de nuevas herramientas digitales a la docencia de las matemáticas es inevitable y se intensificará en los próximos años. El uso de los ordenadores para resolver problemas no implica la desaparición del razonamiento lógico, además, ambas formas de trabajar se complementan y pueden, incluso, potenciarse entre sí. Por tanto, debemos formar a los alumnos en el uso de las nuevas tecnologías para resolver, visualizar o interactuar con un problema matemático y su aplicación en la vida diaria.

Un libro que aborda estas y otras inquietudes relacionadas con las matemáticas y la educación es \cite{B.22}. En él se hace un repaso de distintos ámbitos relacionados con la educación matemática tanto en las etapas obligatorias como en la post-obligatoria, así como cuestiones que deben ser transversales a lo largo de toda la enseñanza de las matemáticas. También se detallan algunos aspectos sobre la formación profesional del profesorado de matemáticas. Por tanto, ha sido una obra de referencia en las dos últimas décadas y puede ser un complemento a la lectura de este artículo.

A lo largo del artículo trataremos los siguientes aspectos:
\begin{enumerate}
    \item En primer lugar, sentaremos el marco normativo, más concretamente, hablaremos sobre las competencias STEM y digital.
    \item Analizaremos el manejo que tienen los profesores y los alumnos en las herramientas digitales relacionadas con las matemáticas. Para ello, se ha llevado a cabo una encuesta a docentes y alumnos sobre el uso de diferentes herramientas como LaTeX, Wolfram Alpha o Geogebra.
    \item Utilizaremos los programas de libre acceso \textit{WxMaxima} y \textit{Geogebra} para resolver un ejercicio de la prueba de acceso a la Universidad de Extremaudra y analizar cómo podrían ayudar al alumnado.
    \item Dada la necesidad de formación en herramientas digitales específicas de matemáticas, propondremos un curso para iniciar a docentes y alumnos en el manejo de WxMaxima y Geogebra.
    \item Finalmente, cerraremos el documento con una sección dedicada a conclusiones obtenidas y posibles líneas de investigación que se podrían explorar a partir de este trabajo.
\end{enumerate}

\section{Competencia STEM y Competencia digital}

La educación de nuestros días se basa en el desarrollo del currículo mediante las competencias básicas o clave (según la ley a la que hagamos referencia). Aunque en el panorama internacional ya se hablaba sobre la educación por competencias desde que Philippe Perrenoud las introdujese en 1998 \cite{P.98} y, posteriormente, la \textit{Recomendación del Parlamento Europeo y del Consejo 2006/962/CE, de 18 de diciembre de 2006, sobre las competencias clave para el aprendizaje} es la primera vez que un organismo europeo recomendaba la implantación de este tipo de educación. En España hicieron su aparición en la \textit{Ley Orgánica 2/2006, de 3 de mayo, de Educación}, aunque no se mencionaban las mismas de forma expresa, ya dejaba ver el carácter europeísta del que se quería dotar a la eduación. Las competencias se fueron asentando con el paso de los años, hasta que a finales de 2013 la \textit{Ley Orgánica 8/2013, de 9 de diciembre, para la mejora de la calidad educativa}, LOMCE en adelante, recoge por primera vez en España las siete competencias claves del sistema educativo español y sienta las relaciones entre las competencias, los contenidos y los criterios de evaluación de los distintos niveles educativos a través de la \textit{Orden ECD/65/2015, de 21 de enero, por la que se describen las relaciones entre las competencias, los contenidos y los criterios de evaluación de la educación primaria, la eduación secundaria obligatoria y el bachillerato}. El caracter competencial en la educación secundaria sigue vigente con la nueva legislación, la \textit{Ley Orgánica 3/2020, de 29 de diciembre, por la que se modifica la Ley Orgánica 2/2006, de 3 de mayo, de Educación} (LOMLOE). En ella se describen las competencias clave, entre las que se encuentran la \textit{competencia Matemática y de ciencias, tecnología e ingeniería (STEM)} y la \textit{competencia digital}.

\begin{figure}[h]
    \centering
    \begin{tikzpicture}
\draw (0,0) node[anchor=east]{$A$}
  -- (0,3) node[anchor=east]{$C$}
  -- (4,0) node[anchor=west]{$B$}
  -- cycle;
\end{tikzpicture}
    \caption{Triángulo que representa una rampa.}
    \label{fig:triangulo}
\end{figure}
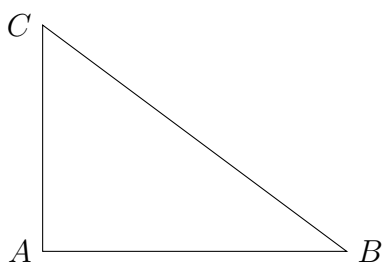

Se establece que la competencia matemática es aquella que ``\textit{implica la capacidad de aplicar el razonamiento matemático y sus herramientas para describir, interpretar y predecir distintos fenómenos}'' \cite[Anexo I,2.a]{OrdenComp}. Por tanto, el estudio de las asignaturas de matemáticas no debe buscar la memorización de fórmulas, problemas típicos o definiciones concretas, sino que, a partir de ciertos conocimientos teóricos, ha de permitir interpretar el mundo que nos rodea para resolver ploblemas con herramientas matemáticas. No se trata entonces de aprender cuáles son las herramientas matemáticas, sino cómo aplicar estas herramientas en el mundo exterior. Por ejemplo, no hay razón ninguna para memorizar la famosa fórmula del Teorema de Pitágoras si no sabemos en qué situaciones es pertinente el uso de este conocimiento matemático. Así, esta fórmula puede servir como mero ejercicio mental de memorización, o puede servirnos para analizar una rampa de un acceso para minusválidos que nos encontremos en cualquier acera de nuestra ciudad. Por ejemplo, si el triángulo de la Figura \ref{fig:triangulo} representa una de dichas rampas 

entonces podemos calcular la longitud de dicha rampa gracias a que $a^2=b^2+c^2$ y también podemos ver el ángulo de inclinación de la rampa con las relaciones trigonométricas y establecer si la rampa cumple o no con las medidas para la accesibilidad del moviliario urbano. Se trata por tanto de relacionar las fórmulas y herramientas matemáticas con el mundo que nos rodea, de forma que lo hagamos entendible desde un punto de vista matemático. De todo ello trata el libro \textit{Mirar la ciudad con ojos matemáticos}, \cite{B.20}, L. Blanco establece distintos itinerarios por la ciudad de Badajoz en los que iremos paseando a la vez que realizamos pequeñas actividades de reflexión o de cálculo matemático en las que emplearemos algunas de las herramientas que aprendemos en la escuela y que muchos alumnos piensan que no tienen utilidad.

En el informe PISA \cite[p. 7]{PISA2021} se introduce el concepto de \textit{alfabetización matemática} que, a su vez es recogido en castellano en \cite[p. 10]{curriculumMat}:
\begin{quotation}\it
    La alfabetización matemática es la capacidad de un individuo de razonar matemáticamente y de formular, emplear e interpretar las matemáticas para resolver problemas en una amplia variedad de contextos de la vida real. Esto incluye conceptos, procedimientos, datos y herramientas para describir, explicar y predecir fenómenos. Ayuda a los individuos a conocer el papel que cumplen las matemáticas en el mundo y hacer los juicios y tomar las decisiones bien fundamentadas que necesitan los ciudadanos reflexivos, constructivos y comprometidos del siglo XXI.
\end{quotation}
Continuando con el mismo documento, el CEMat aporta algunas indicaciones que deben dejar de realizarse en la Educación Secundaria Obligatoria y el Bachillerato, por ejemplo ``utilizar fórmulas dadas como modelos de problemas del mundo real, sin analizar de dónde surge el modelo'' o ``expresar ecuaciones de una función de forma analítica, con el exclusivo objeto de representarlas gráficamente''. Todo ello sugiere que las matemáticas y el mundo que conocemos deben estar relacionados y el objeto de estudio de las matemáticas debe ser precisamente entender dicha relación.

Respecto a la competencia digital \cite[Anexo I, 3]{OrdenComp} ``es aquella que implica el uso creativo, crítico y seguro de las tecnologías de la información y la comunicación''. De forma que 
\begin{quotation}\it
    Esta competencia supone, además de la adecuación a los cambios que introducen las nuevas tecnologías [...] para ser competente en un entorno digital.
    
    Requiere de conocimientos relacionados con el lenguaje específico básico: textual, numérico, icónico, visual, gráfico y sonoro, así como sus pautas de decodificación y transferencia. Esto conlleva el conocimiento de las principales aplicaciones informáticas.
\end{quotation}
Por tanto, con ojos matemáticos, el uso de las nuevas tecnologías puede servirnos para complementar las explicaciones teóricas. Al igual que el uso de la calculadora no implica que la suma de fracciones no tenga que ser estudiada, por ejemplo, el aprendizaje en el uso de herramientas informáticas para el cálculo matemático no implica que esos contenidos teóricos no deban ser explicados, sino que el alumno tendrá una forma más de comprobar si los resultados que va obteniendo son correctos. En el \cite{libroblanco} se pone este mismo ejemplo de la calculadora como herramienta plenamente instaurada en el aula, pero que ``como toda herramienta, es indispensable aprender a utilizarla para que el alumnado no la considere un sustituto del razonamiento matemático'', es decir, no deben verse las herramientas digitales como un sutitutivo del razonamiento, sino como herramientas que exploten ``las posibilidades que ofrece para agilizar procesos de cálculo en las aulas y para que los alumnos puedan centrarse en la comprensión de los nuevos conceptos''.

Según \cite{M.07}, ``La tecnología como recurso de aprendizaje tiene que estar integrada en el currículum, es decir, el maestro formado tiene que saber a la hora de diseñar una actividad si decide emplear un recurso tecnológico, cuándo y cómo hacerlo'' y así se establece en la Orden ECD/65/2015. En ella podemos apreciar que se indica textualmente que se debe conocer el lenguaje básico de las herramientas digitales en relación con conocimientos numéricos y gráficos, lo que implica el conocimiento de las principales aplicaciones informáticas. Parece, por tanto, necesario relacionar las dos competencias con las que estamos tratanto, la matemática y la digital, de manera que se complementen mutuamente. A pesar de que el uso de herramientas digitales está expresamente recogido en el currículo de la educación secundaria de las asignaturas de matemáticas, el informe realizado por la Real Sociedad Matemática Espaóla y la Sociedad Científica Informática de España en 2020 \cite{matesinfor} reconoce que ``la gran mayoría de los libros sigue incorporando la tecnología de modo anecdótico y muy ligada a la calculadora'', además de reflejar la gran diferencia que hay en los docentes, ``desde profesorado muy tecnologizado que incorpora geometría dinámica, hojas de cálculo, software de representación gráfica y hasta lenguajes de programación en su docencia de matemática, hasta profesorado que prohíbe el uso de la calculadora en ESO''. En \cite{ER.14}, se propuso una serie de actividades con Geogebra basados en rompecabezas, llegando a la conclusión de que ``aprender Geometría utilizando rompecabezas fue percibido por la mayoría de los estudiantes como una experiencia agradable que les permite descubrir nuevas cosas'' y que al docente le sirve ``no sólo para estimular el pensamiento geométrico, sino también fomentar el interés y motivación hacia aprendizaje de la Geometría''. Además, el uso de las nuevas tecnologías en el aula tiene importantes implicaciones pedagógicas, como pueden ser, según \cite{M.07}: ``la creación un entorno interactivo de enseñanza / aprendizaje en el cual los aprendices pueden ser indistintamente emisores y receptores de información, lo que provoca alta motivación en el aprendiz'', que repercutirá directamente en el clima de la clase; ``la enseñanza se adapta a las necesidades específicas del alumno'', lo que favorece una de las obligaciones de todo docente, la atención a la diversidad; ``se facilita la enseñanza cooperativa'', favoreciendo así las relaciones sociales y la empatía entre discentes; y por último ``se fomenta el autoaprendizaje'', es decir, estamos repercutiendo positivamente en la adquisición de otra de las competencias básicas de especial importancia en la educación integral de los alumnos, \textit{aprender a aprender}.

Una vez que hemos visto las dos competencias fundamentales que se buscan estudiar en este trabajo, cabe destacar que la Real Sociedad Matemática Española está haciendo importantes esfuerzos por estudiar cómo aunar estas dos competencias durante todo el período de escolaridad. Más concretamente, en \cite{libroblanco} se establece que ``la educación y, específicamente, el aprendizaje de las matemáticas, no pueden ser ajenas a los cambios que nos conducen hacia una sociedad más tecnificada y digitalizada''. Analizaremos en la próxima sección, la necesidad de formación del profesorado en el uso y utilidad de las herramientas digitales. Según Carioca et. al. \cite{montanero}, la necesidad de formación del profesorado en el uso de las herramientas digitales es una necesidad prioritaria de nuestros días, no tanto en el uso de las propias herramientas, sino en la utilidad didáctica que tienen y en cómo llevarlas al aula. Además, Martín destaca en \cite[p. 44]{M.07} que ``debido a la rapidez de los avances tecnológicos se hace más patente que nunca la asunción del profesorado en general de la necesidad de un reciclaje en su formación tecnológica y en el uso pedagógico de la tecnología''. Necesidad que se subraya en el Libro Blanco de las Matemáticas \cite[p.33]{libroblanco} más de una década después: ``el uso de tecnologías en las aulas requiere de un conocimiento didáctico que va más allá del conocimiento de su simple uso''.

\section{Las herramientas digitales en docentes y alumnos}

\subsection{Objetivo y justificación}

Ya hemos señalado en las secciones anteriores, el papel de las nuevas tecnologías en la enseñanza de las matemáticas. Sin embargo, aún hay docentes que se limitan al uso de las clases magistrales para la enseñanza de esta materia. El objetivo principal de este artículo es realizar un estudio en docentes para comprobar el uso que le dan a distintas herramientas digitales, así como el grado de conocimiento que poseen de ellas. Más aún, nos interesa ver cómo percibe el alumnado la implementación de las herramientas digitales en la enseñanza de las matemáticas. Todo esto nos lleva a realizar una encuesta a docentes y alumnado, en la que se les pregunte por distintos aspectos relacionados con las herramientas digitales y, más específicamente, por aquellas relacionadas con las matemáticas. A raíz de estas encuestas, podremos establecer distintas comparativas en cuanto al uso de las herramientas digitales por parte de los docentes en el aula o por parte de los alumnos en sus casas, así como intentar entender los beneficios y los perjuicios que declaran ambos colectivos en el uso de las herramientas digitales en la enseñanza de las matemáticas.

En esta sección veremos las principales respuestas a las encuestas realizadas a docentes y alumnos durante el curso 2020-2021. La muestra tomada para el estudio en los docentes ha sido el conjunto de profesores de Ciencias (Matemáticas, Física, Química y Biología) del centro en el que el autor realizó las prácticas del Máster de Formación del Profesorado, durante el curso 2020-2021, el Colegio Santa Teresa de Jesús en Badajoz, cuya titularidad es privada--concertada para los niveles de la ESO y privada--no concertada para los niveles de Bachillerato. Además, para obtener más información, se ha ampliado la muestra con el conjunto de los profesores de Ciencias del Instituto de Educación Secundaria ``El Sur'', de la ciudad de Lepe, cuya titularidad es pública. La muestra tomada en los alumnos ha sido exclusivamente del Colegio Santa Teresa de Badajoz, más concretamente los alumnos de los grupos donde mi tutora de prácticas imparte clases, es decir, 4.º de la ESO, 1.º y 2.º de Bachillerato, tanto la rama de Matemáticas de Ciencias Sociales como de la de Matemáticas para Ciencias. Por supuesto, la encuesta ha sido anónima y voluntaria.

Las preguntas que se hicieron a ambos colectivos pueden encontrarse en los siguientes enlaces:

Docentes: \\ \url{https://drive.google.com/file/d/1r-_YReEOhFgaTqtXX2TqPu94N4EQoDFW/view?usp=share_link}

Alumnado: \\ \url{https://drive.google.com/file/d/1460CqUUTl63VheDltHetF_r2OuMZmJeC/view?usp=share_link}

Se pueden observar que las preguntas son parecidas, de forma que podamos establecer algunas relaciones. Analizaremos primero las respuestas de los docentes, después la del alumnado, y finalmente comparemos ambos colectivos.

\subsection{Análisis en docentes}

Entre los docentes se han registrado un total de 12 respuestas. Empezando por la formación, observamos que la mayoría ($58.3\%$) de docentes provienen de licenciaturas, por lo que han pasado al menos diez años desde que egregaron y, posiblemente, no tengan toda la formación en el uso de las TIC con fines didácticos como sería deseable. Sin embargo, todos los docentes reconocen su utilidad y sólo el $8.3\%$ admite no saber usarlas.

Algunas de las respuestas más interesantes a la pregunta \textit{¿Cómo crees que las herramientas digitales (en general) pueden ayudar a los alumnos en el aula o fuera de ella?}, han sido:\textit{
\begin{itemize}
    \item Pueden visualizar contenidos complejos de reproducir sin digitalización.
    \item La motivación hacia la asignatura es mayor utilizando una herramienta digital.
    \item Pueden ayudar a comprender y aprender más y mejor que con el método tradicional.
    \item Les ayuda a que los estudios sean más cercanos.
\end{itemize}}

Además, dos tercios del profesorado encuestado afirma haber realizado cursos de formación en herramientas digitales en el último año, quizá debido a que han tenido que actualizarse rápidamente por los confinamientos y la situación del COVID--19, mientras que un $16.7\%$ afirma haber realizado cursos en los últimos cinco años y el otro $16.7\%$ afirma no haber realizado cursos desde hace más de diez años. Muchos de los cursos ofertados por el Centro de Profesores y Recursos (CPR) son de manejo elemental y, precisamente por esto, algunos de los docentes que usan estas herramientas terminan no haciendo los cursos. 

\begin{figure}[h]
    \centering
    \includegraphics[scale=0.6]{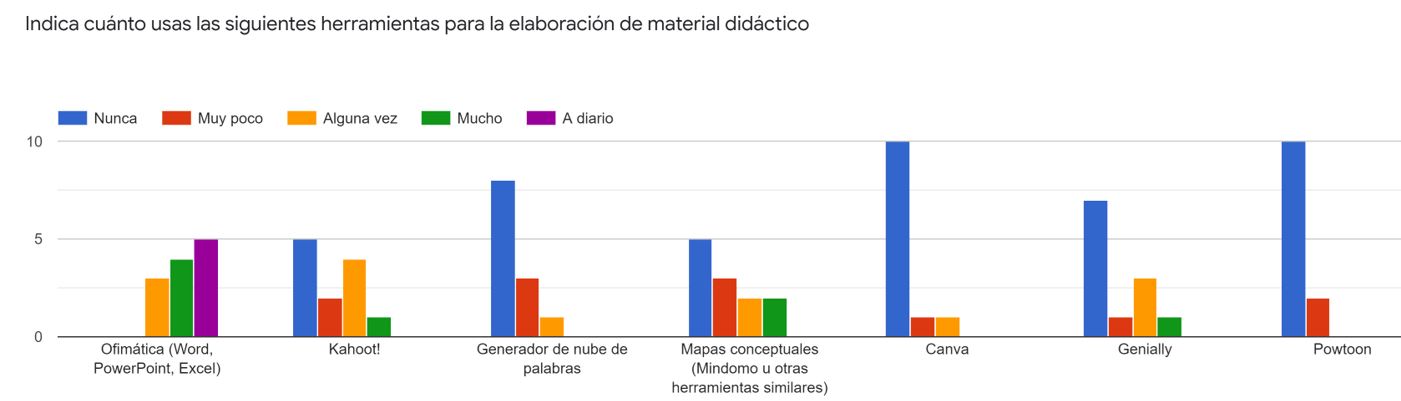}
    \caption{Uso de herramientas para elaborar material didáctico en docentes.}
    \label{respdoc:5}
\end{figure}

Se preguntó también sobre el nivel de uso de distintas herramientas (ofimática, \textit{Kahoot!}, generador de nubes de palabras, mapas conceptuales, \textit{Canva}, \textit{Genially} y \textit{Powtoon}) para la elaboración de material didáctico. Las respuestas se recogen en la Figura \ref{respdoc:5}.

Cabe destacar el uso casi a diario de las herramientas de ofimática, y un uso algo menor para \textit{Kahoot!} y herramientas de creación de mapas conceptuales. El resto de aplicaciones tiene un uso prácticamente nulo.

\begin{figure}[h]
    \centering
    \includegraphics[scale=0.42]{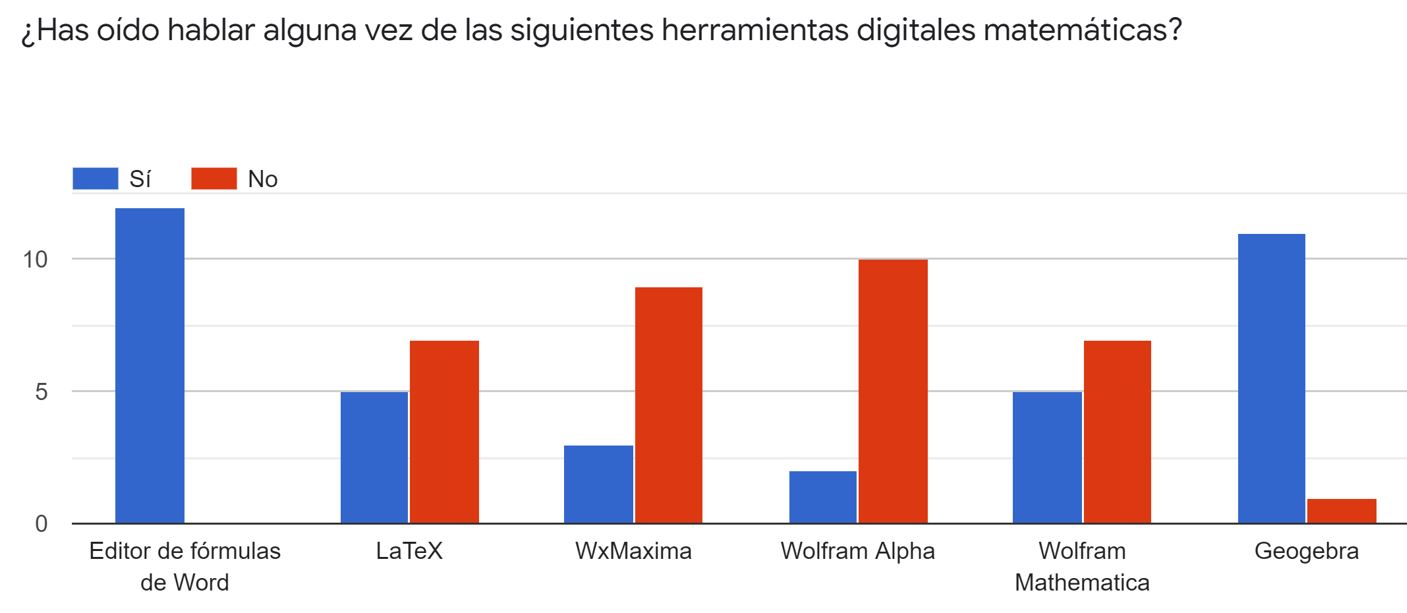}
    \caption{Conocimiento de herramientas digitales específicas en docentes.}
    \label{respdoc:7}
\end{figure}

En el ámbito específico de las herramientas digitales relacionadas con las matemáticas, sobre el conocimiento en general de ciertas herramientas se han obtenido las respuestas recogidas en la Figura \ref{respdoc:7}.

Donde apreciamos que todos los docentes conocen el editor de fórmulas de Word, pero que la mayoría (siete de los doce encuestados) no conocen LaTeX para la edición de texto científico. Sobre las herramientas de cálculo destaca que casi todos conocen Geogebra, pero la mayoría desconocen programas de cálculo simbólico como WxMaxima, Wolfram Alpha o Wolfram Mathematica. Centrándonos, por tanto, en el uso del editor de fórmulas de Word, cinco docentes afirman usarlo con mucha frecuencia y dos de ellos a diario, indicando, en la siguiente pregunta, que han aprendido a usarlo, mayoritariamente de manera autodidacta. En cuanto a Geogebra, la mayoría de respuesta se ubica en que su uso es esporádico (alguna vez) y que la mayoría, de nuevo, han sido autodidactas en su aprendizaje.

Algunas opiniones de los docentes sobre cómo ayudan estas herramientas específicamente matemáticas a los alumnos son:\textit{
\begin{itemize}
    \item Permite obtener multitud de ejemplos y generalizar contenidos trabajados. Muy útil en geometría, estadística y análisis.
    \item Pueden hacer más atractiva la asignatura y les puede motivar más su estudio.
    \item Facilitar los cálculos, comprobar sus resultados, visualizar.
    \item Les hacen desarrollar la 'imaginación' y capacidad de deducción y resolución de problemas y teorías.
\end{itemize}}

Esto se corresponde con la idea generalizada que ya puntualizamos en la sección anterior y que en \cite{curriculumMat} transmite el CEMAT, y es que es necesario ``usar utilidades informáticas para desarrollar estructuras conceptuales'' en detrimento del cálculo con ``lápiz y papel'' ya que ``solucionar problemas con ayuda del ordenador es un ejercicio que permite adquirir la costumbre de enfrentarse a problemas predefinidos de una forma rigurosa y sistemática''.

Como último apunte, algunas de las respuestas más interesantes de los docentes a si les gustaría recibir algún tipo de formación sobre el uso de estas herramientas fueron:\textit{
\begin{itemize}
    \item Geogebra, sobre la realización de recursos interactivos.
    \item Geogebra y WxMaxima.
    \item Sí, sería muy bueno aprender todas las funcionalidades de todas estas herramientas.
    \item Sí, sobre Geogebra, con un curso práctico.
    \item Sí. Me gustaría saber emplear en mis clases más de las herramientas digitales que ya uso.
\end{itemize}}
De estas respuestas se deduce el interés del profesorado en la formación en herramientas digitales específicas de las matemáticas para su utilización en el aula y que, además opinan que pueden tener un importante beneficio en el alumnado. Así, en la sección \ref{sec:curso} se propondrá un curso formativo para docentes en las herramientas WxMaxima y Geogebra.

\subsection{Análisis en alumnado}

En este caso, hemos obtenido 40 respuestas por parte del alumnado. Empezaremos localizando en los cursos a estos alumnos, el $20\%$ corresponde a alumnos de 4.º de la ESO, el $52.5\%$ a alumnos de 1.º de Bachillerato y el $27.5\%$ de 2.º de Bachillerato. Para empezar, cabe destacar que todos afirmaron que las herramientas digitales son útiles, y sólo el $5\%$ de ellos admitió no saber utilizarlas.

Algunas de las respuestas más interesantes a la pregunta \textit{¿Cómo crees que las herramientas digitales (en general) pueden ayudarte en el aula o fuera de ella?}, han sido:\textit{
\begin{itemize}
    \item Sobre todo en la dinámica que estableces con el profesor, teniendo en cuenta que es mucho más entretenida la clase y se te pasa mucho más rápido.
    \item A la hora de poder hacer trabajos, estudiar o incluso informarme de cosas que no sabía antes que pueden ser interesantes.
    \item Creo que la comunicación mejoraría mucho entre alumnos y profesores. También haría que las clases fueran más dinámicas.
    \item Ayudan a organizarte mejor, y a veces te hacen las cosas más fáciles, más visuales. Además te permite preguntar al profesor aunque no esté en horario de clases.
\end{itemize}
}

\begin{figure}[h!]
    \centering
    \includegraphics[scale=0.65]{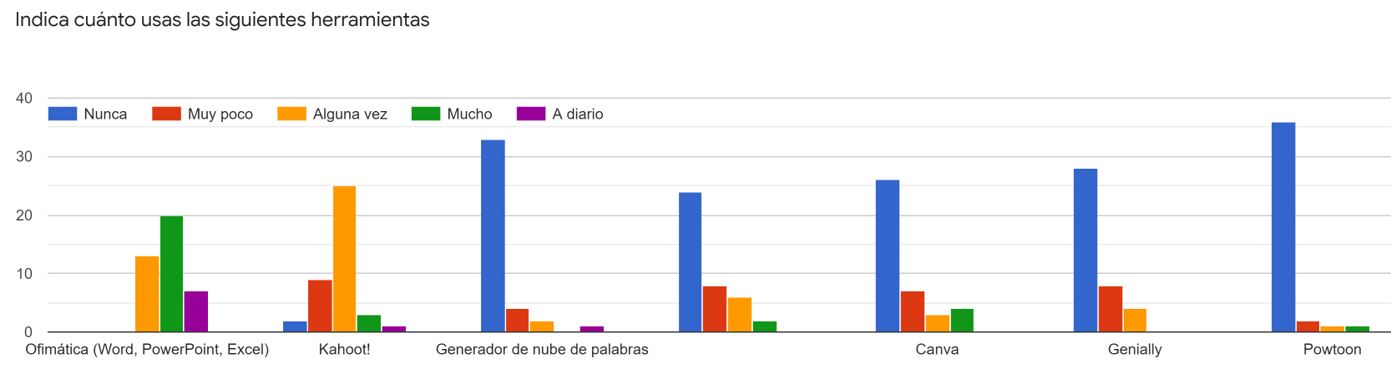}
    \caption{Uso de herramientas digitales generales en el alumnado.}
    \label{respalu:2}
\end{figure}

Cuando se pregunta al alumnado sobre el uso de distintas herramientas generales (ofimática, \textit{Kahoot!}, generador de nubes de palabras, mapas conceptuales, \textit{Canva}, \textit{Genially} y \textit{Powtoon}), se obtienen las respuestas recogidas en la Figura \ref{respalu:2}.

Donde destaca especialmente el uso bastante extendido de las aplicaciones de ofimática y algo menor el uso de \textit{Kahoot!}, además de un uso prácticamente nulo del resto de herramientas.

\begin{figure}[h]
    \centering
    \includegraphics[scale=0.42]{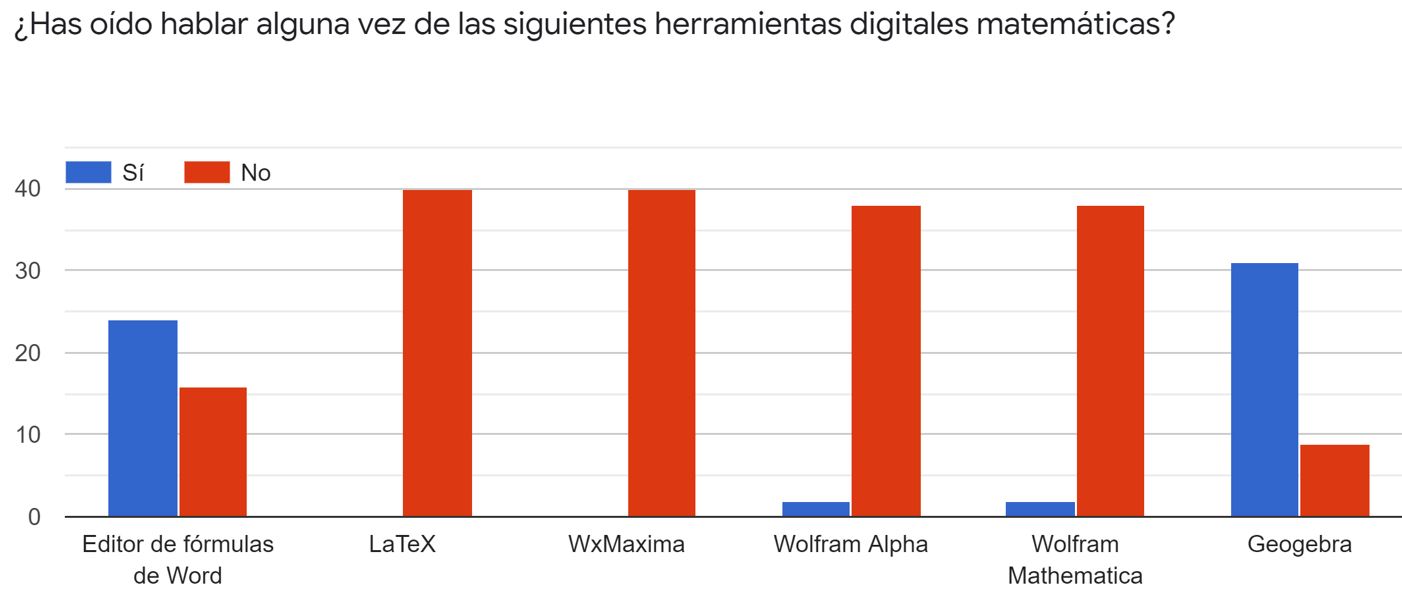}
    \caption{Conocimiento de herramientas digitales específicas en el alumnado.}
    \label{respalu:3}
\end{figure}

Centrándonos en las herramientas digitales matemáticas, se preguntó a los alumnos por las mismas herramientas que a los docentes, obteniendo las respuestas recogidas en la Figura \ref{respalu:3}.

\begin{figure}[h]
    \centering
    \includegraphics[scale=0.6]{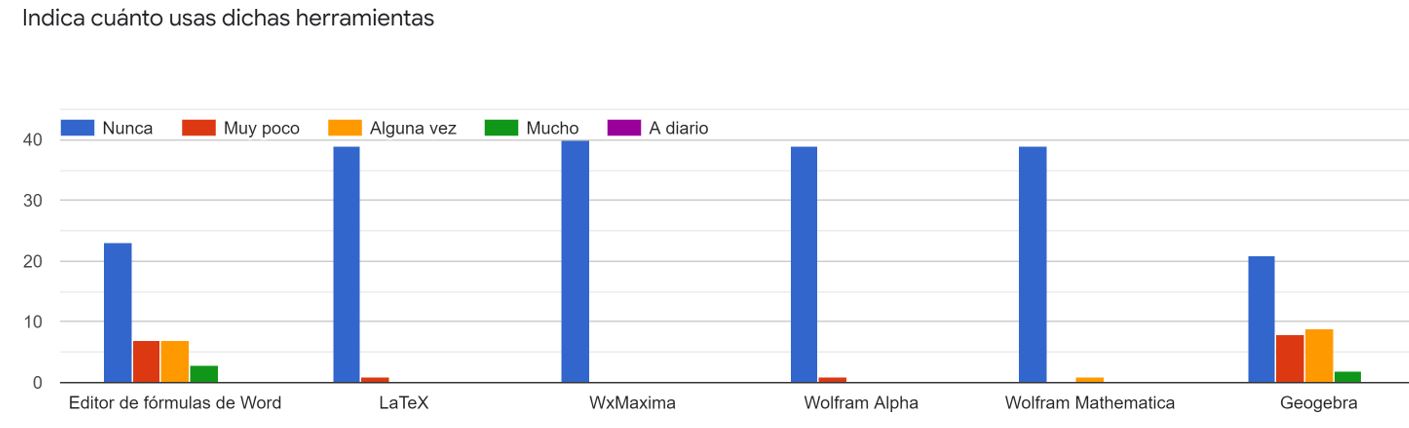}
    \caption{Uso de herramientas digitales específicas por el alumnado.}
    \label{respalu:5}
\end{figure}

Siguen destacando el editor de fórmulas de Word y Geogebra, aunque sus usos sean prácticamente nulos tal y como se recoge en la Figura \ref{respalu:5}.

Así, algunas de las respuestas más interesantes a la pregunta \textit{¿Te gustaría recibir algún tipo de formación sobre alguna de las herramientas matemáticas?}, fueron: \textit{
\begin{itemize}
    \item Sí, sinceramente de todas. Porque recientemente he visto como era la de Geogebra y, me ha parecido que está muy avanzada y tiene muy buenas cosas.
    \item No tengo ninguna preferencia y la formación debería ser una breve introducción o algo por el estilo.
    \item Sí, como WxMaxima u otra para saber más sobre ellas. A través de algún seminario o vídeos explicativos.
    \item No me interesa en este momento.
\end{itemize}
}

Observamos que, a pesar de haber algunas respuestas negativas a recibir algún tipo de formación, la mayoría de alumnos reconoció querer formarse en este tipo de herramientas. Y, finalmente, sobre por qué creen que estas les pueden ayudar en casa o en el aula, algunas de las respuestas más representativas fueron:\textit{
\begin{itemize}
    \item Una mayor precisión en casos de gráficas, exactitud en cálculos y control y organización cuando se necesario manejar mucho cálculos diferentes.
    \item A la hora de las correcciones son más precisas que a lo mejor hacer una representación gráfica en la pizarra.
    \item Haciendo ejercicios que no entiendo, para poder prepararme un examen y así saber si hago bien los ejercicios.
    \item No sé a los demás, pero a mi durante la cuarentena me salvaron la evaluación, porque ponía los ejercicios y ya está ja ja (Pero era porque no me enteraba, lo prometo).
\end{itemize}
}
Por tanto, a pesar de que en la respuesta anterior hubo alumnos que no querían formarse, en esta pregunta se aprecia una opinión generalizada en torno a que el uso de este tipo de herramientas puede ayudar en casa y en el aula. Incluso la última respuesta la he incluido porque hay alumnos para los que el uso de las herramientas informáticas es circunstancial y anecdótico, y no está relacionado con el trabajo o el aprendizaje matemático.

\subsection{Análisis de las diferencias entre docentes y alumnos}

Teniendo en cuenta lo expuesto en las secciones de análisis de las respuestas de docentes y alumnos, podemos destacar las siguientes conclusiones importantes:
\begin{itemize}
    \item Ambos colectivos consideran útil el uso de las herramientas digitales.
    \item De manera general, docentes y discentes tienen conocimientos sobre herramientas poco creativas y efectivas, como pueden ser las de ofimática; empieza a haber discrepancias en herramientas que necesitan un mayor grado de creatividad como pueden ser \textit{Kahoot!}, \textit{Canva} o \textit{Genially}.
    \item El uso de dichas herramientas ofimáticas está muy extendido, prácticamente a diario por parte de docentes y alumnos, pero el resto de herramientas son usadas de manera esporádica.
    \item Ambos grupos consideran que las redes sociales pueden ayudar a la comunicación docente--discente y por tanto resaltan la importancia de una conversación fluida y, en cierto modo, distendida.
    \item Sobre las herramientas de carácter matemático, el desconocimiento es prácticamente absoluto, salvando quizá los casos del editor de fórmulas de Word y Geogebra. El conocimiento del primero de ellos quizá sea debido al amplio uso que han reconocido de las herramientas de ofimática. El conocimiento y uso de Geogebra quizá merezca un estudio más en profundidad, pero en cualquier caso es una de las herramientas más sencillas e intuitivas de utilizar y que tiene una dilatada experiencia en el sector educativo, por lo que muchos docentes, cada vez más, la utilizan en clase con distintos fines. Por ejemplo \cite{GST.14}, estudia el uso de Geogebra como facilitador del aprendizaje en el Análisis, concretamente  establece que Geogebra favorece ``la utilización de las funciones de variable real como instrumento de modelización y herramienta de solución de situaciones problemáticas intra y extra matemáticas'', así como ``la visualización dinámica de los comportamientos funcionales''.
    \item Aunque algunas herramientas, tanto generales como matemáticas, son conocidas, la gran mayoría de herramientas son, o bien desconocidas, o bien de poca utilización. Esto nos lleva a preguntarnos si quizá se deba a que no han recibido ningún tipo de formación sobre sus diferentes usos. Ya que, en cierta medida \textit{no podemos utilizar algo que no conocemos}. En esta línea se investigó en el artículo \cite{montanero} la formación continuada en herramientas informáticas del profesorado de Extremadura y del Alentejo portugués, determinando que ``en Extremadura [...] la formación permanente del profesorado no necesita incidir tanto en la preparación técnica en el campo de la informática cuanto en la utilización didáctica del ordenador en el contexto escolar''. Es decir, tenemos profesores que entienden el uso del ordenador y de herramientas informáticas, pero no saben cómo implementarlas en el aula.
\end{itemize}

Así, a modo de conclusión, observamos que hay consenso en que las nuevas tencnologías están aquí para quedarse, por lo que, en nuestro caso particular, docentes y discentes entienden la importancia y el amplio abanico que pueden ofrecer herramientas informáticas a la mejora de la calidad educativa.

\section{WxMaxima y Geogebra como herramientas didácticas}

El software \textit{WxMaxima} es un programa de software libre, esto es, gratuito, lo cuál facilita su difusión y utilización, a diferencia de otros programas que requieren una licencia de pago. Puede descargarse para distintos sistemas operativos como Windows, Mac OS, Linux o Ubuntu, por lo que prácticamente podremos instalarlo en cualquier ordenador e incluso tiene su versión en Android, por lo que algunos alumnos podrán incluso trabajar con sus dispositivos móviles. Se trata de un programa que, más allá del propio cálculo, tiene grandes implicaciones pedagógicas, como apunta Jorquera en \cite[p.11-12]{J.08}:
\begin{quotation}\it
    Permite realizar cálculos reales, de mayor dificultad matemática evitando perder tiempo en el cálculo rutinario, con lo que se puede dedicar más tiempo a la explicación de los conceptos que a las habilidades de cálculo.
    
    Será útil en algunos casos, e inútil en otros. No hay que forzar su uso en todas las ocasiones ya que esto sería un error, pero hay contextos en los cuales el error sería no usarlo.
    
    Se fomenta más el trabajo creativo en detrimento del rutinario.
\end{quotation}

Por su parte, Geogebra tiene una versión de escritorio y una versión en línea, ambas gratuitas. Además, en la página de inicio de Geogebra existen un gran abanico de actividades en línea y descargables para realizar en el aula con esta aplicación. Está en español, lo cual es una gran ventaja para muchos docentes y alumnos. Y tiene una interfaz sencilla e intuitiva, que cualquiera puede aprender a manejarla simplemente probando sus distintas herramientas. A pesar de ello, tiene un gran número de funciones más complejas para crear actividades muy interesantes. En este sentido, ``Geogebra mejora significativamente el aprendizaje de la capacidad de razonamiento y demostración, de comunicación matemática así como la capacidad de resolución de problemas'' \cite{tesis}, así como ``contribuyó a que mejorasen sus actitudes hacia las matemáticas durante su uso, exhibiendo gusto, implicación y autoconfianza en matemáticas'' \cite{tesis}.

Veamos, mediante un ejercicio de la prueba de acceso a la Universidad de Extremadura correspondiente al bloque de Geometría de la asignatura Matemáticas II, cómo WxMaxima y Geogebra se complementan para una resolución sencilla. Para ello, resolveremos el ejercicio manualmente y con ambos programas, argumentando cuál es más idoneo en cada situación.

Este ejercicio apareció en la convocatoria ordinaria del curso 2020--2021 y su encunciado fue el siguiente:
\begin{quote}
    Sea el plano $\Pi$ de ecuación $2x+y-z-2=0$ y la recta $r$ dada por $\frac{x}{3}=\frac{y-2}{-3}=\frac{z-1}{3}$.
    \begin{enumerate}
        \item Estudie la posición relativa de la recta respecto al plano.
        \item Calcule la distancia de la recta al plano.
    \end{enumerate}
\end{quote}

\subsubsection*{Resolución manual}
Debemos tener en cuenta que un punto de la recta es $P=(0,2,1)$ y su vector director es $\vec{u}=(3,-3,3)$. Por otro lado el vector normal del plano es $\vec{n}=(2,1,-1)$. Así, como
$$
\vec{u}\cdot\vec{n}=3\cdot2-3\cdot1+3\cdot(-1)=6-3-3=0
$$
y comprobando si $P$ pertenece, o no, a $\Pi$,
$$
2\cdot0+2-1-2=-1\neq0.
$$
Luego el plano y la recta son paralelos. La distancia vendrá dada por
$$
\textup{dist}(P,\Pi)=\frac{|2\cdot 0+1\cdot 2-1\cdot 1-2|}{|(2,1,-1)|}=\frac{1}{\sqrt{6}}.
$$

\subsubsection*{Resolución con WxMaxima}
La idea que tenemos que seguir es la misma que la resolución manual. En primer lugar, definimos los vectores director de la recta y normal del plano, y realizamos su producto escalar (Figura \ref{resolucion1:maxima2}):

\begin{figure}[h]
    \centering
    \subfloat{\includegraphics[scale=0.65]{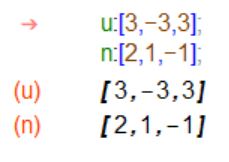}}
    \subfloat{\includegraphics[scale=0.65]{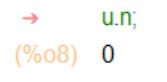}}
    \caption{Definición de los vectores y producto escalar de ambos.}
    \label{resolucion1:maxima2}
\end{figure}

Se define la función que nos indica si un punto pertenece o no al plano $\Pi$ y sustituimos el punto $P=(0,2,1)$ (Figura \ref{resolucion1:maxima3}):

\begin{figure}[h]
    \centering
    \subfloat{\includegraphics[scale=0.65]{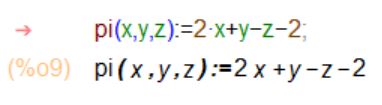}}
    \subfloat{\includegraphics[scale=0.65]{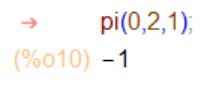}}
    \caption{Definición del plano $\Pi$ y comprobación de si el punto $P$ pertenece a $\Pi$.}
    \label{resolucion1:maxima3}
\end{figure}

\begin{figure}[h]
    \centering
    \includegraphics[scale=0.65]{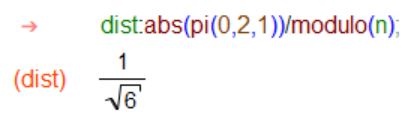}
    \caption{Cálculo de la distancia de $r$ a $\Pi$.}
    \label{resolucion1:maxima4}
\end{figure}

Como $\vec{u}\cdot\vec{n}=0$ y $P\notin\Pi$, se deduce que la recta $r$ y el plano $\Pi$ son paralelos, por lo que su distancia puede ser calculada como sigue (Figura \ref{resolucion1:maxima4})

Y el ejercicio quedaría resuelto.
\newpage
\subsubsection*{Resolución con Geogebra}

\begin{figure}[h]
    \centering
    \includegraphics[scale=0.37]{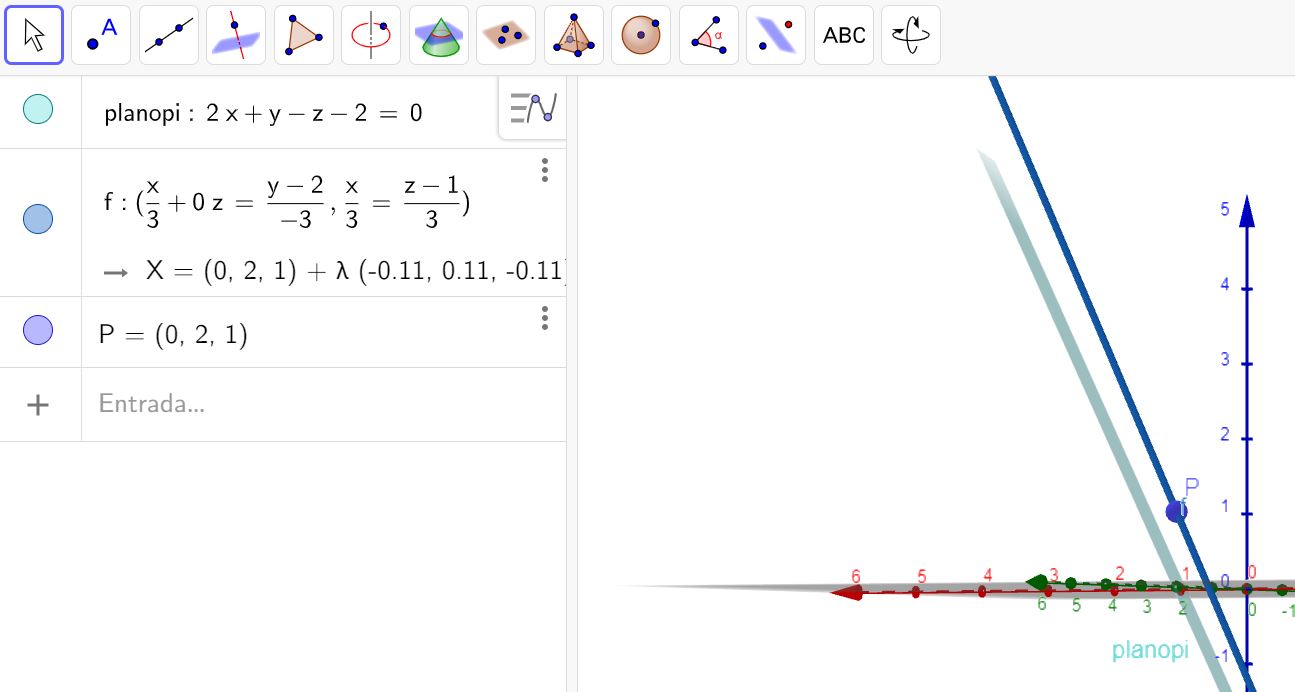}
    \caption{Definición de $\Pi$, $r$ y $P$.}
    \label{resolucion1:geo1}
\end{figure}

Primero definimos el plano y la recta dados, así como el punto $P=(0,2,1)$ (Figura \ref{resolucion1:geo1}).

\begin{figure}[h]
    \centering
    \includegraphics[scale=0.36]{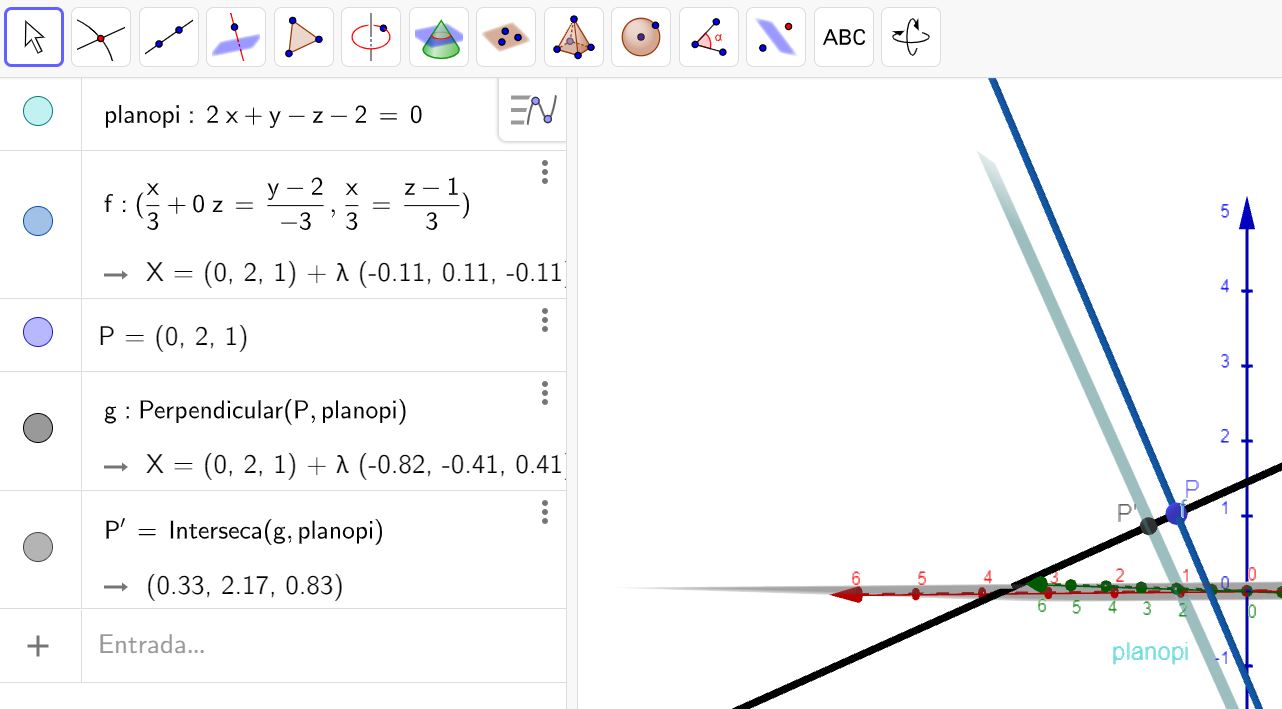}
    \caption{Recta perpendicular a $\Pi$ y proyección de $P$ en $\Pi$.}
    \label{resolucion1:geo2}
\end{figure}

A continuación, trazamos la recta perpendicular al plano que pasa por $P$ y señalamos como $P'$ el punto donde intersecta esta perpendicular con el plano $\Pi$, es decir, la proyección de $P$ en $\Pi$ (Figura \ref{resolucion1:geo2})

\begin{figure}[h]
    \centering
    \includegraphics[scale=0.36]{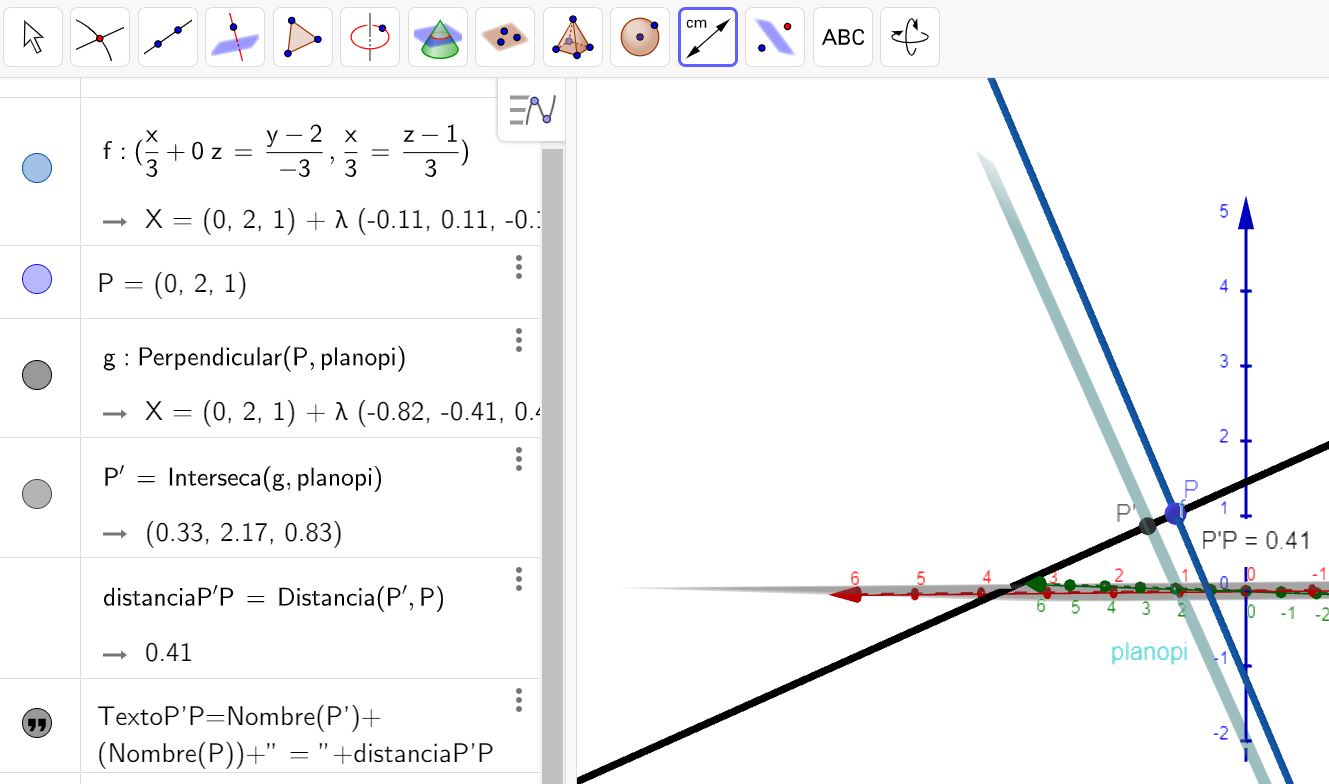}
    \caption{Distancia de $P$ a $P'$.}
    \label{resolucion1:geo3}
\end{figure}

Para finalizar, sólo tendremos que utilizar la herramienta de distancia señalando los puntos $P$ y $P'$ (Figura \ref{resolucion1:geo3}).

Obteniendo el resultado aproximado de $0'41\approx\frac{1}{\sqrt{6}}$, concluyendo así el ejercicio.

\subsubsection*{Comparativa entre las resoluciones}
En este caso es clara la diferencia entre las tres resoluciones:
\begin{itemize}
    \item La resolución manual es tediosa y falta de visualización, algo clave en la parte de Geometría, ya que es la parte de las matemáticas más visual.
    \item La resolución con WxMaxima hace sencillos los cálculos, pero de nuevo carece de visualización. Bien es cierto que, además, tiene un lenguaje particular y es necesario saber escribir las operaciones y funciones, ya que, de lo contrario, podemos tardar mucho en resolver un ejercicio.
    \item La resolución con Geogebra es eminentemente visual, de hecho, para resolver el primer apartado basta con representar ambos objetos para darnos cuenta de su posición relativa. Más complicado es, sin embargo, el segundo apartado, ya que debemos conocer que la distancia entre una recta y un plano paralelos es la distancia entre un punto cualquiera de la recta hasta su proyección en el plano, lo cual no es tan sencillo de realizar en Geogebra ya que tendremos que calcular la perpendicular, la intersección y después la distancia. Además, observamos que la distancia obtenida es una aproximación y no el valor exacto.
\end{itemize}
Por tanto, mi conclusión en este ejercicio es que la solución óptima vendría dada por una mezcla de los dos programas, el primer apartado se resolvería fácilmente con Geogebra mientras que para el segundo apartado emplearía WxMaxima.

\vspace*{-0.5cm}\section{Curso de formación para el profesorado}\label{sec:curso}

\vspace*{-0.2cm}Teniendo en cuenta que a lo largo del trabajo se ha pretendido inculcar una visión de las herramientas digitales como facilitadoras del aprendizaje, pero que tanto docentes como alumnos no conocen su funcionamiento, el autor propone un curso/seminario formativo orientado a docentes. Estará compuesto por cuatro sesiones y se trabajarán los aspectos y utilidades fundamentales de WxMaxima y Geogebra recogidos en las Tablas \ref{tab:ses1}, \ref{tab:ses2}, \ref{tab:ses3} y \ref{tab:ses4}.

\vspace*{-0.2cm}

\renewcommand{\tablename}{Tabla}
\begin{table}[h]
\begin{center}
\begin{tabular}{ | m{2.5cm} | m{12.5cm} | }
\hline Sesión: & Primera \\ \hline
Título: & Introducción a WxMaxima I: Álgebra lineal \\ \hline
Contenidos y videotutoriales & \begin{enumerate}\small
    \item Instalación y entorno principal: \url{https://youtu.be/O6ibBL20vi8}
    \item Consideraciones generales: \url{https://youtu.be/THKn0QJkq4A}
    \item Matrices y operaciones elementales: \url{https://youtu.be/EX_6vDxvPi0} y \url{https://youtu.be/kiWPhfywUQM}
    \item Problemas usuales que se pueden tratar: \url{https://youtu.be/3aNCG-AXMls}
\end{enumerate} \\ \hline
\end{tabular}\vspace*{-0.2cm}
    \caption{Contenidos de la primera sesión.}
    \label{tab:ses1}
\end{center}
\end{table}

\vspace*{-1.3cm}

\begin{table}[h]
\begin{center}
\begin{tabular}{ | m{2.5cm} | m{12.5cm} | }
\hline Sesión: & Segunda \\ \hline
Título: & Introducción a WxMaxima II: Análisis y cálculo vectorial. \\ \hline
Contenidos y videotutoriales & \begin{enumerate}\small
    \item Definición y representación de funciones: \url{https://youtu.be/Nhq9lnEEJpo}
    \item Límite de funciones: \url{https://youtu.be/_ZDJqtwE7aE}
    \item Derivadas e integrales: \url{https://youtu.be/-C5Q4d2ChD8}
    \item Raíces, máximos y mínimos: \url{https://youtu.be/6wB7r3Vaa3c}
    \item Definición de vectores: \url{https://youtu.be/gsCpgfacn-E}
    \item Producto escalar, vectorial y mixto: \url{https://youtu.be/_7F2Xnlc_V8}
    \item Ángulo entre vectores y posiciones relativas: \url{https://youtu.be/CQxS0ui6o58} y \url{https://youtu.be/g5NHlJ5B3VM}
\end{enumerate} \\ \hline
\end{tabular}\vspace*{-0.2cm}
    \caption{Contenidos de la segunda sesión.}
    \label{tab:ses2}
\end{center}
\end{table}

\vspace{-1cm}

\begin{table}[h]
\begin{center}
\begin{tabular}{ | m{2.5cm} | m{12.5cm} | }
\hline Sesión: & Tercera \\ \hline
Título: & Introducción a Geogebra I: 2D. \\ \hline
Contenidos y videotutoriales & \begin{enumerate}\small
    \item Instalación y entorno principal: \url{https://youtu.be/4Zc1PZdIE3o}
    \item Definición de puntos y rectas: \url{https://youtu.be/Xtm_IUjU3vo}
    \item Operaciones con rectas: \url{https://youtu.be/kCibwIYRxMc}
    \item Representación de funciones: \url{https://youtu.be/nbSjilpgVcg}
\end{enumerate} \\ \hline
\end{tabular}
    \caption{Contenidos de la tercera sesión.}
    \label{tab:ses3}
\end{center}
\end{table}

\vspace*{-2.4cm}

\begin{table}[h!]
\begin{center}
\begin{tabular}{ | m{2.5cm} | m{12.5cm} | }
\hline Sesión: & Cuarta \\ \hline
Título: & Introducción a Geogebra II: 3D. \\ \hline
Contenidos y videotutoriales & \begin{enumerate}\small
    \item Algunas consideraciones sobre el entorno 3D: \url{https://youtu.be/sftFwEx3JAM}
    \item Definición de puntos, rectas y planos: \url{https://youtu.be/OTagU5OawSA}
    \item Operaciones con rectas y planos: \url{https://youtu.be/eeGZ7hWtVH0}
    \item Distancias y ángulos entre rectas y planos: \url{https://youtu.be/rth9TvXsbRM}
\end{enumerate} \\ \hline
\end{tabular}
    \caption{Contenidos de la cuarta sesión.}
    \label{tab:ses4}
\end{center}
\end{table}

\newpage
En total, este curso/seminario consta de 21 vídeotutoriales con una duración de dos horas y media aproximadamente. Trece de ellos están dedicados, entre las dos primeras sesiones, a WxMaxima, con una duración de 1 hora y 25 minutos aproximadamente; y ocho vídeos, los correspondientes a las sesiones tercera y cuarta, dedicados a Geogebra, con una duración aproximada de una hora y cinco minutos.

Como hemos venido haciendo hincapié, su objetivo es brindar a los docentes las herramientas y utilidades básicas de estos dos programas para que puedan adaptar los ejercicios del día a día en el aula e implementar el uso de las herramientas digitales para la resolución y visualización de contenidos matemáticos, así como de la mejora de la asimilación de los conceptos y una mejora de la predisposicióna a aprender matemáticas.

\section{Conclusiones y líneas de investigación}

A modo de conclusión, podemos extraer algunas de las ideas fundamentales que se han desarrollado a lo largo del trabajo:

\begin{itemize}
    \item En primer lugar, el aprendizaje del contenido matemático y de procesos de resolución de problemas no está reñido con el hecho de aprender el uso de herramientas digitales. Esto se debe a que ambos métodos de resolución se complementan. Hemos visto, dentro de los ejercicios resueltos, que un programa puede servir mejor que el otro para ``hacer cuentas'' o para representar, pero en cualquiera de los dos programas, se ha de tener una visión general de la parte teórica para poder seguir el hilo argumental del problema y obtener la solución. Esto puede ayudar a la autocorrección de los ejercicios que, habitualmente, los alumnos tienen como tarea para casa; también puede ayudar a que alumnos con altas capacidades indaguen e investiguen en mayor profundidad, buscando nuevos retos y problemas más complicados; e incluso, puede ayudar, al aprendizaje por descubrimiento o al \textit{just in time teaching}, es decir, darle respuestas al alumno en el momento en que se haga las cuestiones, por lo que los programas pueden servir para dejarles probar y, como docentes, les ayudaremos en sus procesos de descubrimiento.
    \item En segundo lugar, creo que hay mucho camino por delante en cuanto a la implantación de las nuevas tecnologías en los centros educativos. Aunque es verdad que cada vez se invierte más en adaptar las aulas a las nuevas tecnologías (pizarras digitales, proyectores o aulas virtuales), no podemos decir lo mismo sobre la inversión en la formación de los docentes en el uso de estas herramientas. Esto implica que las nuevas tecnologías pueden ser en ciertos casos, vistas como una fuente adicional de estrés en el profesorado, por ejemplo ante la posibilidad de que el docente tenga preparada una clase y, el día que pretenda impartirla, no funcione internet; o que las herramientas sean vistas más como un elemento extraño que como un medio educativo. Tampoco los docentes conocen todas las funcionalidades que estas les ofrecen, por tanto, muchos de ellos seguirán con sus rutinas sin saber los beneficios de dichas herramientas.
    \item Como hemos dicho, hay mucho camino por recorrer, lo cual es bueno, prácticamente cualquier iniciativa formativa será bien acogida por la comunidad educativa, tanto de formación al profesorado como de formación al alumnado, ya que ambos colectivos observan que las herramientas digitales tienen un gran potencial, pero no saben cómo aprovecharlo.
\end{itemize}

Por último, teniendo en cuenta estas conclusiones, podemos proponer las siguientes líneas de investigación para las que será necesario un análisis más detallado de las diferencias en el uso y en la concepción de las herramientas digitales en la educación secundaria:

\begin{itemize}
    \item Lo primero en lo que deberíamos pensar es en realizar una encuesta a un espacio muestral más amplio y elegido considerando ciertas variables (formación inicial del profesorado, titulación de acceso al cuerpo de profesores de secundaria, años de experiencia como docente, contexto socioeconómico del centro, etc.), con el objetivo de poder obtener conclusiones más significativas para una población mayor.
    \item Una vez realizado el curso introductorio de las herramientas digitales WxMaxima y Geogebra, realizar un seguimiento, por ejemplo, tomar dos grupos de alumnos que cursen la misma asignatura de matemáticas en el mismo curso. A uno de estos grupos se le asignará un profesor que no haya realizado el curso introductorio y seguirá una metodología sin TIC. Al otro grupo, se le asignará un profesor que sí haya realizado el curso, que haya intentado introducir estas herramientas a lo largo de las explicaciones. Finalmente, se comparará si el uso de las herramientas digitales ha influido no sólo en los resultados, sino en la aceptación de la asignatura de matemáticas, en la motivación con que van a clase y en la satisfacción y la percepción de aprendizaje de los alumnos.
    \item Por último, en caso de que lo anterior refleje un aumento significativo de los resultados académicos y la motivación del alumno por la asignatura, cabría plantear un curso de material más avanzado con el que puedan seguir profundizando en el uso de WxMaxima y Geogebra los docentes que hicieron el primer curso.     Así, de nuevo, tras un dominio más avanzado de las herramientas podríamos hacer un estudio comparativo entre ambos grupos, uno con un docente que tenga un menor dominio, que solo haya realizado el primer curso, y otro con un profesor que haya realizado el curso avanzado.
\end{itemize}

\nocite{LOE,LOMCE,LOMLOE,RecEurComp}

    \bibliographystyle{plain}
	\bibliography{HerramientasDigitalesSecundaria-plano.bib}
\end{document}